\documentclass[preprint2]{emulateapj}

\usepackage{natbib}  
\usepackage{amssymb}

\usepackage{comment}
\usepackage{graphicx}

\bibpunct{(}{)}{;}{a}{}{,} 

\def\cm{\,{\rm cm}}

\def\ergscm2 {erg\,s$^{-1}$cm$^{-2}$}

\def\cm2 {cm$^{-2}$}

\def\aap {A\&A}

\shorttitle{The  ``double-humped" super-luminous supernova SN2006oz}
\shortauthors{Rachid Ouyed and Denis Leahy}

\begin{document}

\title{The peculiar case of the ``double-humped" super-luminous supernova SN2006oz}

\author{Rachid Ouyed  and Denis Leahy }
\affil{Department of Physics and Astronomy, University of Calgary, 
2500 University Drive NW, Calgary, Alberta, T2N 1N4 Canada}

\begin{abstract}
SN2006oz is a super-luminous supernova with a mysterious bright precursor that has 
  resisted explanation in standard models. However,  such a precursor has been predicted
   in the dual-shock quark nova (dsQN) model of super-luminous supernovae -- the precursor
   is the SN event while the main light curve of the SLSN is powered by the Quark-Nova (QN;
   explosive transition of the neutron star to a quark star). 
   As the SN is fading, the QN  re-energizes the SN ejecta,  producing  a ``double-humped" light curve. 
   In this paper,  we show the dsQN model successfully reproduces the observed light curve of SN2006oz.
\end{abstract}

\section{Introduction}

Supernova (SN) 2006oz  (Leloudas et al. 2012) is a newly-recognized member of the class of H-poor, super-luminous supernovae (i.e. SN2005ap-like; Quimby et al. 2011). 
The bolometric light curve shows a precursor ``plateau" with a duration between 6-10 days in the rest-frame and it is followed
 by a dip, after which the luminosity begins to rise.  
 This subsequent rise was fit using three different models (see Chatzopolos et al. 2011): (i) input from radioactive decay;
(ii) a magnetar spin-down model; (iii)  a circum-stellar matter (CSM) interaction. The Nickel decay
model is least likely since it requires unreasonable amounts (10.8$M_{\odot}$) of $^{56}$Ni  in a total ejecta mass of 14.4$M_{\odot}$. 
 In addition the SN was not detected nine months later, inconsistent with the standard decay curve for $^{60}$Co. The magnetar
  and CSM models present a decent but not accurate fit to the data (see Figure 7 Leloudas et al. 2012). 
   In general, to explain SN 2005ap-like objects  (Chomuik et al. 2011) the suggested models
    require rather extreme additional conditions. The magnetar model requires
   initial spin periods near break-up (1-2 ms) while CSM interaction models require expelling
   several solar masses of H-poor material in the few years before the explosion: this  has never
   been observed from W-R stars (see Chomuik et al. 2011 for details).

 Existing Models for the precursor (Dessart et al. 2011) are too dim to explain it. The only explanation offered by Leloudas et al. (2012)
 was a recombination wave in oxygen around the progenitor star with no physical cause for the wave suggested. 
 At the current stage, none of the above models can account   for the precursor.
  This begs for other alternatives which can explain the precursor and the main peak   of SN2006oz  self-consistently.

  The energy in the precursor we estimate to be $\sim 10^{49}\ {\rm erg}\times t_{\rm pre, 10}$ where
   $t_{\rm pre,10}$ is the duration of the precursor in units of 10 days (limited by the observations from about 7 days
   to 12 days).  We note that this energy is typical of  brighter Type-II SNe (e.g. Young 2004) suggesting that the precursor  could in fact
   be the SN explosion proper. This would require that the main peak has a separate physical origin. 
    The quark nova (QN) was proposed as an alternative explanation for SN 2006gy and other
   SLSNe including SN 2005ap (Leahy \& Ouyed 2008).  In Ouyed et al. (2009a), we also emphasize the
   lightcurve of the preceding SN, giving a ``double-humped" lightcurve very much reminiscent
   of that of SN2006oz.

In this paper we focus on studying the lightcurve of SN2006oz in the context of our model: the dual-shock QN (dsQN) model.
The paper is organized as follows: in \S 2 we give a brief review of the dsQN model.
In \S 3 we show that the main peak  and the precursor of SN2006oz are self-consistently fit
 by the dsQN.   We briefly conclude in \S 4.

\section{Our model}

The quark nova (QN) was proposed as an alternative explanation for SN 2006gy (Leahy \& Ouyed 2008; Ouyed et al. 2009a). A QN is expected to occur when the core density of a neutron star reaches the quark de-confinement density and triggers a violent (Ouyed et al. 2002) conversion to the more stable strange quark matter (Itoh 1970; Bodmer 1971; Witten 1984). The novel proposition was made, that during the spin-down evolution of the neutron star, a detonative (Ouyed et al. 2002; Niebergal et al. 2010) phase transition to up-down-strange triplets would eject the outer layers of the neutron star at ultra-relativistic velocities (Ker\"anen et al. 2005; Ouyed \& Leahy 2009). Studies of neutrino and photon emission processes during the QN (Vogt et al. 2004; Ouyed et al. 2005) have shown that these outermost layers (of $\sim 10^{-4}$-$10^{-3} M_{\odot}$ in mass) can be ejected with up to $10^{53}$ erg in kinetic energy.

If the time delay ($t_{\rm delay}$) between SN and QN explosions is lengthy the SN ejecta will have dissipated such that the QN essentially erupts in isolation. However, when $t_{\rm delay}$ is on the order of days to weeks,  a violent collision occurs reheating the extended SN ejecta (Leahy \& Ouyed 2008; Ouyed et al. 2009a). The emission from the re-shocked SN ejecta declines as the photosphere recedes, eventually revealing a mixture of the
SN and QN material with unique chemical signatures (Jaikumar et al. 2007;  Ouyed et al. 2009a; Ouyed et al. 2010; Ouyed et al. 2011).

The basic physical processes involved in our model are: (i) a SN explosion at time $t=0$ with homologously
expanding ejecta with the outermost velocity at $v_{\rm SN}$;
(ii) a QN explosion at time $t_{\rm delay}$ which launches a  shock at velocity $v_{\rm QN}$  into the 
preceding SN ejecta. This second shock reheats the SN ejecta to $T_{\rm QN}$; (iii) The QN shock breaks out from
 the SN ejecta at time  $t_{\rm delay} + t_{\rm prop}$, where $t_{\rm prop}$ is the 
 time for the QN shock to propagate through the SN ejecta.  The reheated SN ejecta expands while
  radiating and undergoing adiabatic expansion losses. We approximate the evolution
   of the photosphere using photon diffusion in a pure Thompson scattering medium (see Leahy\&Ouyed 2008).
   A key feature of this model is that the shock reheating occurs at large radius (because of the time delay) so that
    standard adiabatic losses inherent to SN ejecta are far smaller. In effect the SN provides the material
     at large radius and the QN re-energizes it giving the large luminosity compared to a normal SN.

\section{Application to SN2006oz}

Figure 1 shows the observed SN2006oz light curve  from Leloudas et al. (2011; their Table 3).
We use the g-band data which has the best time coverage and also lowest errors for most times.
The data is plotted in days at the source using the measured redshift of $z\sim 0.376$. 
We converted apparent g-band magnitudes to absolute g-band magnitudes using the 
corresponding luminosity distance for the standard model (Wright 2006).
 We converted the suggested extinction correction (B-V) from  Leloudas et al. (2011)
  to (g-V) and included it, even though it was small. 
  
  Our model also agrees with the early and late  upper limits from Leloudas et al. (2011) although they are not
  plotted here because we chose  to show better  the firm detections.  For the
  SN lightcurve (i.e. the first hump), we prefer to compare to an observed light curve.
 We use the light curve of SN1999em from Bersten\&Hamuy (2009)
  which has good time coverage in the first 50 days. 
  Bersten et al. (2011) fitted hydrodynamic models to SN1999em and derived a progenitor mass of $19M_{\odot}$ (similar
  in mass to the SN progenitor we used in our QN model),
  radius of $800R_{\odot}$, explosion energy of $1.25\times 10^{51}$ erg and 
  $^{56}$Ni mass of 0.056$M_{\odot}$. This gave a luminosity  at $5$ days of
  $10^{42.4}$ erg s$^{-1}$. We scaled the  bolometric magnitude by +2 to  
   represent a more energetic SN.  This is not unreasonable  
    since the range in brightness of Type II SNe
  varies  considerably with many models giving brighter SN than 1993em (e.g. Young 2004).

In the QN model the progenitor initial mass is in the range of 20-40$M_{\odot}$ (Leahy\&Ouyed 2009; Ouyed et al. 2009b; Ouyed et al. 2010)
to create a massive neutron star with core density near the instability to convert to quark matter (Niebergal et al. 2010).
This motivates our choice of SN  ejected mass of 20$M_{\odot}$. Best fits from our previous studies
of SLSNe yielded time delays of $\sim 10$ days which motivates the time delays that
we explored. For SN2006oz the shown fit (see Figure 1) uses $t_{\rm delay}=6.5$ days,
$v_{\rm QN}= 5000$ km s$^{-1}$ and a preceding SN ejecta with an average velocity of $v_{\rm SN}\simeq 1900$ km s$^{-1}$.
The combined light from the SN and from the QN-reheated SN ejecta give a reasonable
 fit to the observations with a self-consistent model.

\begin{figure}
\begin{center}
\includegraphics[width=0.46\textwidth]{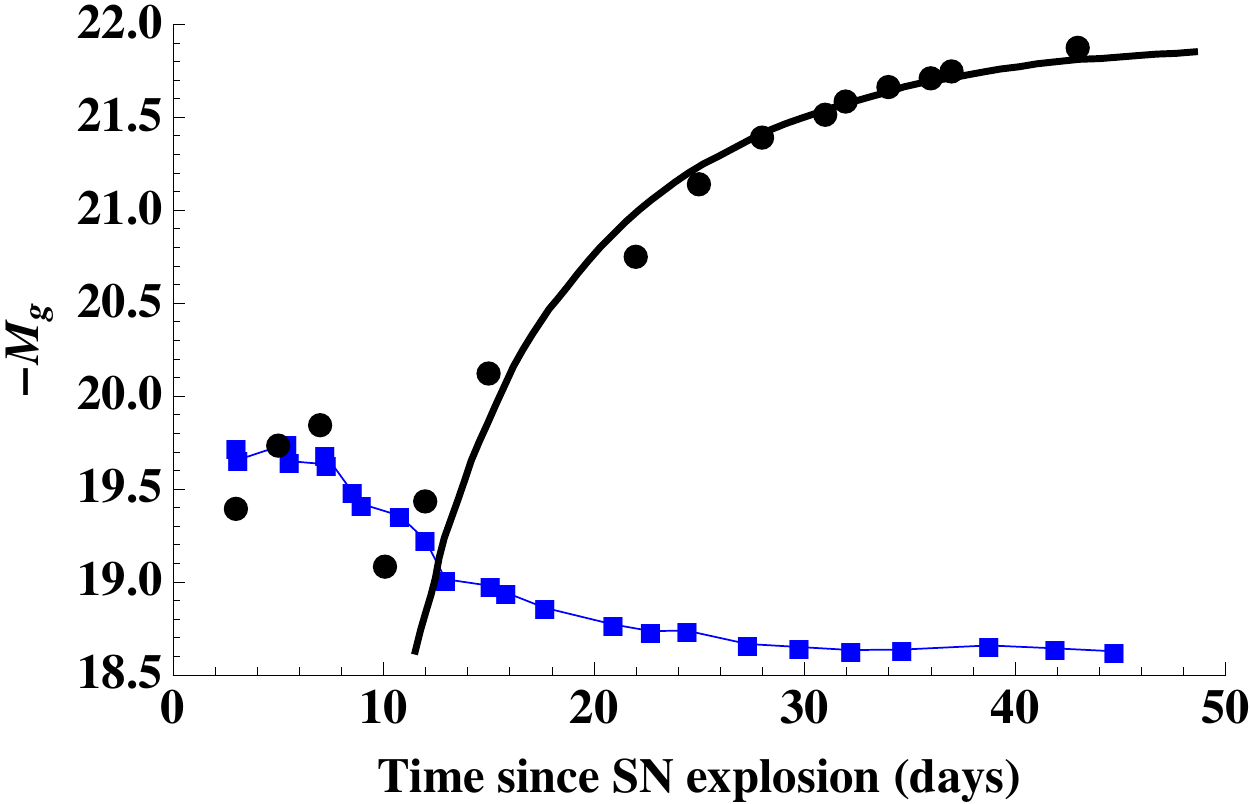}
\caption{SN2006oz g-band absolute magnitude light curve (solid circles). The dsQN model is
calculated for $M_{\rm ejecta}=20M_{\odot}$ and 
$t_{\rm delay}= 6.5$ days (see text for other parameters). The proto-type SN light curve (connected squares)
is a scaled version of that  observed for SN1999em (see text).}
\end{center}
\end{figure}

\section{Discussion and Conclusion}

Recent observations (such as the Texas SN search; Quimby et al. 2005
 and the Catalina Real-Time Transit Survey; Drake et al. 2009) have revealed a new class of supernovae,
the SLSNe and among these are the SN2005ap-like (H-poor) SLSNe.
 These events have proven challenging to explain. SN2006oz was the first
 to have clearly shown a bright precursor with absolute magnitude of $\sim$ -19 to -20.
  We suggest this precursor is a type II SN and the main event is the QN (i.e.
   the SN envelope re-heated by the QN).

 We note from Leloudas et al. (2012) the intriguing possibility of an intrinsic
precursor event in SN 2005ap-like objects. In our model, there must be a
normal SN ($ -20 <  M_{\rm bol} <  -15 $) preceding the SLSN  if the delay is long enough
that the SN light curve is not buried in the QN one. The precursor SN 
should be detectable  in sensitive and early enough observations of  SN 2005ap-like explosions.

  SN 2005ap-like objects occur at a rate of $< 1/ 10^4$ core-collapse SNe (Quimby et al. 2011).
dsQNe are expected to be rare:   The QNe rate is estimated to be $\sim 1/1000$ core-collapse events 
 with 1/10 of them having time delays in the appropriate range to produce dsQNe ($t_{\rm delay}\sim $ 5-30 days;
  Staff et al. 2006; Jaikumar et al. 2007; Leahy\&Ouyed 2008; Leahy\&Ouyed 2009; Ouyed et al. 2009b).
  This order of magnitude estimate is consistent with the rate of SN 2005ap-like events.

  Our model applies to both H-rich and H-poor SLSNe -- the key ingredient is
a progenitor in the right mass range to produce a massive enough NS but
not a black hole.  
 We note that in both cases, the QN shock reheats the SN envelope so H-poor/H-rich
 progenitors would give H-poor/H-rich spectra. In this context, we expect
 H-poor SLSNe to occur in higher-metallicity environment (i.e. higher stellar mass loss-rates).
  Low-metallicity progenitors would lose less mass and would more likely be H-rich and
 should in principle have more massive envelopes.

Upcoming observations from the large SN surveys should reveal more SLSNe
 and more of these with precursors. In our model, these precursors are
  type II SNe which should be verifiable with good enough photometry
  and/or spectroscopy.  In addition, the overall shape of the 
  SLSN lightcurve should vary from a single hump to a double hump
  depending on the time delay between the SN and the QN explosions.

\begin{acknowledgements}
This research  is supported by  operating grants from the
National Science and Engineering Research Council of Canada (NSERC). 
\end{acknowledgements}


\begin{thebibliography}{99}


\bibitem[]{} Bersten, M.~C., \& Hamuy, M.\ 2009, ApJ, 701, 200

\bibitem[]{} Bersten, M.~C., Benvenuto, O., \& Hamuy, M.\ 2011 [arXiv:1101.0467v2]

\bibitem[]{}   Bodmer, A. R. 1971, Phys. Rev. D, 4, 1601

\bibitem[]{}   Chomiuk, L., Chornock, 
R., Soderberg, A.~M., et al.\ 2011, ApJ, 743, 114

\bibitem[]{}  Drake, A. J., Djorgovski, S. G., Mahabal, A., et al. 2009, ApJ, 696, 870

\bibitem[]{}  Dessart, L., Hillier, D. J., Livne, E., et al. 2011, MNRAS, 414, 2985

\bibitem[]{}   Itoh, N. 1970, Prog. Theor. Phys. 44, 291

\bibitem[]{}   Jaikumar, P., Meyer, B.~S., Otsuki, K., \& Ouyed, R.\ 2007, Astronomy\&Astrophysics, 471, 227

\bibitem[]{}   Ker{\"a}nen, P., Ouyed, R., and 
Jaikumar, P.\ 2005, ApJ, 618, 485

\bibitem[]{}   Leahy, D., and Ouyed, R.\ 2008, Mon. Not. Roy. Ast. Soc., 387, 1193

\bibitem[]{}   Leahy, D., \& Ouyed, R.\ 2009, Advances in Astronomy, 2009

\bibitem[]{}   Leloudas, G., 
Chatzopoulos, E., Dilday, B., et al.\ 2012, arXiv:1201.5393

\bibitem[]{}   Niebergal, B., Ouyed, R.,
 \& Jaikumar, P. 2010b, Phys. Rev. C 82, 062801 [arXiv:1008.4806 [nucl-th]]
 
 
\bibitem[]{}  Ouyed, R., Dey, J., \& Dey, M.\ 2002, A\&A, 390, L39

\bibitem[]{}   Ouyed, R., Rapp, R., \& Vogt, C.,
2005, ApJ, 632, 1001


\bibitem[]{}   Ouyed, R., \& Leahy, D.\ 2009, ApJ, 696, 562

\bibitem[]{}   Ouyed, R., Leahy, D., 
\& Jaikumar, P.\ 2009a, Proceedings for "Compact stars in the QCD phase diagram II (CSQCD II)", May 20-24, 2009, KIAA at Peking University, Beijing- P. R. China, [http://vega.bac.pku.edu.cn/rxxu/csqcd.htm],  [arXiv:0911.5424]

\bibitem[]{}   Ouyed, R., Pudritz, 
R.~E., \& Jaikumar, P.\ 2009b, ApJ, 702, 1575

\bibitem[]{}   Ouyed, R., Kostka, M.,  Koning, N., Leahy, D., \& Steffen, W.\ 2010, submitted to MNRAS,  [arXiv:1010.5530]

\bibitem[]{}   Ouyed, R., Leahy, D., 
Ouyed, A., \& Jaikumar, P.\ 2011, Physical Review Letters, 107, 151103


\bibitem[]{}    Quimby, R. M., Castro, F., Gerardy, C. L., et al. 2005, in BAAS, Vol. 37,
American Astronomical Society Meeting Abstracts, 1431

\bibitem[]{}   Quimby, R.~M., et al.\ 
2011, Nature, 474, 487

\bibitem[]{}   Staff, J.~E., Ouyed, R., 
\& Jaikumar, P.\ 2006, ApJ, 645, L145

\bibitem[]{}  Vogt, C., Rapp, R., \& Ouyed, R. 2004,
Nuc. Phys. A, 735, 543

\bibitem[]{}   Witten, E. 1984, Phys. Rev. D, 30,  272

\bibitem[]{}  Wright, E.~L.\ 2006, The Publications of the Astronomical Society of the Pacific, 
118, 1711

\bibitem[]{}   Young, T.~R.\ 2004, ApJ, 617, 
1233 


\end{thebibliography}
\end{document}